\documentclass[letterpaper, 10 pt, conference]{ieeeconf}  

\IEEEoverridecommandlockouts                              

\usepackage{graphicx}
\usepackage{graphics}

\title{\LARGE \bf
Emotion Classification in Response to Tactile Enhanced Multimedia using Frequency Domain Features of Brain Signals
}

\author{Aasim Raheel$^{1}$, Muhammad Majid$^{1}$, Syed Muhammad Anwar$^{2}$, Ulas Bagci$^{2}$ 
\thanks{$^{1}$Aasim Raheel and Muhammad Majid are with Signal, Image, Multimedia Processing and LEarning (SIMPLE) Research Group, Department of Computer Engineering, University of Engineering and Technology, Taxila, 47050 Pakistan
        }%
\thanks{$^{2}$Syed Muhammad Anwar (s.anwar@knights.ucf.edu) and Ulas Bagci are with the Center for Research in Computer Vision (CRCV), University of Central Florida, Orlando, Florida, USA}
        }

\begin{document}

\maketitle
\thispagestyle{empty}
\pagestyle{empty}

\begin{abstract}
Tactile enhanced multimedia is generated by synchronizing traditional multimedia clips, to generate hot and cold air effect, with an electric heater and a fan. This objective is to give viewers a more realistic and immersing feel of the multimedia content. The response to this enhanced multimedia content (mulsemedia) is evaluated in terms of the appreciation/emotion by using human brain signals. We observe and record  electroencephalography (EEG) data using a commercially available four channel MUSE headband. A total of $21$ participants voluntarily participated in this study for EEG recordings. We extract frequency domain features from five different bands of each EEG channel. Four emotions namely: happy, relaxed, sad, and angry are classified using a support vector machine in response to the tactile enhanced multimedia. An increased accuracy of $76.19$\% is achieved when compared to $63.41$\% by using the time domain features. Our results show that the selected frequency domain features could be better suited for emotion classification in mulsemedia studies.

\end{abstract}

\section{INTRODUCTION}

The primary focus of bio-inspired multimedia is to generate and evaluate an immersing environment in terms of appreciation/emotion. Emotions play an important role in content selection and consumption \cite{multimodal}. They are used to evaluate the human experience, while viewing and absorbing the multimedia content. There are six basic types of human emotions including fear, happy, disgust, joy, sad, and anger. Each type of emotion corresponds to a different value for valence and arousal of an individual. Emotions are evoked as a response to an event, where different events/stimuli evokes different types of human emotions. Emotions play a vital role in evaluating various types of multimedia content such as audio, video, and images. They are also helpful in identifying interest of individuals in such content.

\begin{figure*}[!t]
\begin{center}
  \includegraphics[width=160mm]{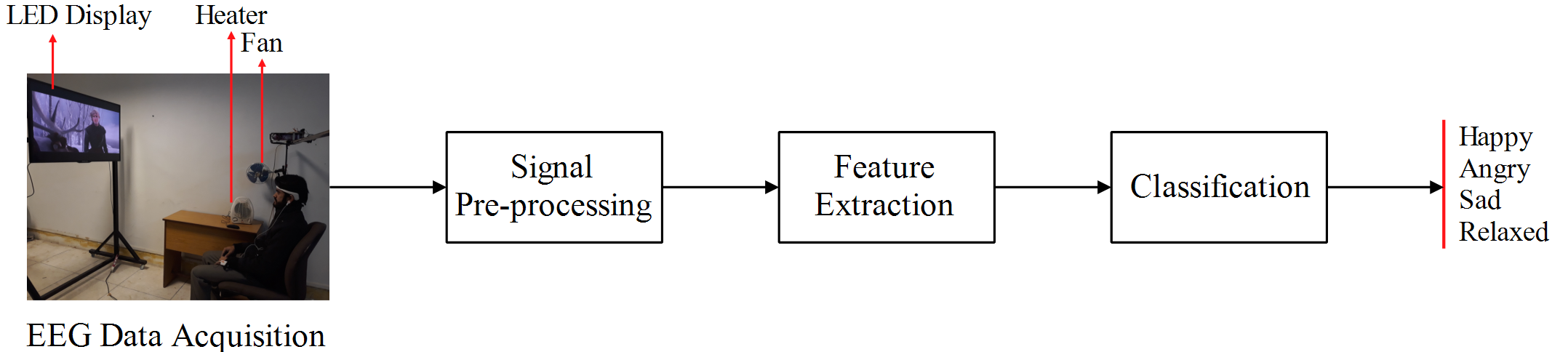}
  
\caption{Proposed methodology for EEG based emotion recognition system by using TEM as a stimuli.}
\label{fig:methodfig}
\end{center}
\end{figure*}

Human senses play a vital role in communication and perception. A higher level of information perception can be achieved by engaging multiple human senses \cite{olfsync2,airflow2}. In the emerging field of human computer interaction, the advancement in various technologies could benefit from the involvement of multiple human senses. This could translate in to better communication with computer applications. The traditional multimedia content engages two senses of a viewer i.e., vision and auditory. Humans are equipped with multiple sensory perception that include senses other than vision and auditory. Hence, the human experience can be enhanced by delivering multimedia content in a way, where multiple sensory organs are engaged. Such a mechanism has been referred to as mulsemedia (multiple sensorial media) in recent literature \cite{mulse}. External devices like scent dispenser, fan, heater, and haptic vest have been used to engage multiple human senses, when combined with traditional multimedia. A fan or a heater is synchronized to engage additional tactile sense within the traditional multimedia content. Different human emotions are evoked by engaging multiple human senses, while viewing certain digital content. Such content can be evaluated by estimating a participant's response in terms of evoked emotions at the time of observation by using electroencephalography (EEG) signals \cite{mytactile}.

Different techniques for emotion recognition have been proposed in literature using EEG by either individually engaging the sense of vision, auditory, olfaction, or tactile or by combining both vision and auditory senses. One of the common indicators of an emotional state, when using EEG is asymmetry of powers on the alpha band which is derived as differences between a symmetric pair of electrodes on the brain \cite{biomedical7}. Other spectral changes and activation in different brain regions are also related to emotions i.e., difference of power on the alpha and theta bands and asymmetry in beta band \cite{biomedical10} at the right parietal region \cite{biomedical10}. Emotion assessment is a subjective phenomenon and different stimuli evokes different emotions. In \cite{bhatti,music2018}, frequency and time domain features were extracted to classify emotions using EEG by using audio music as a stimuli that engaged only the auditory sense of the listener. In \cite{realtimehappiness}, images and audio music were used as a stimuli, thereby engaging only one sense at a time. The emotions have been classified using EEG by extracting frequency domain features. The brain response to tactile sense has been investigated using a time-dependent frequency analysis \cite{brainpleasanttouch}. In \cite{survey108,multimodal}, two human senses were engaged by displaying video clips and emotions were classified using EEG by extracting time domain features.

Recently, tactile enhanced multimedia (TEM) content has been generated by synchronizing video clips with a heater and a fan to engage additional tactile sense of the observer. Such content allows simultaneously engaging three human senses i.e., auditory, tactile, and vision sense. It has been established that the traditional multimedia and its tactile enhanced version are two different stimulus in terms of evoking emotions, based on the statistical analysis of the valence and arousal scores \cite{mytactile}. Four emotions namely: happy, relaxed, sad, and angry have been recognized using statistical time domain features. In our proposed study, frequency domain features were evaluated for emotion recognition using EEG signals in response to the tactile enhanced multimedia content. The length of the feature vector was reduced using frequency domain features when compared to the time domain features.

The rest of the paper is organized as follows. The proposed methodology for emotion recognition is presented in Section \ref{mm}. Experimental results are shown in Section \ref{er} followed by conclusion in Section \ref{cc}.

\section{PROPOSED METHODOLOGY}
\label{mm}
A four step approach was used for EEG based emotion recognition in response to TEM as shown in Fig. \ref{fig:methodfig} and are described in the following subsections.

\subsection{Data Acquisition}
\begin{figure}[b]
\begin{center}
  \includegraphics[width=90mm]{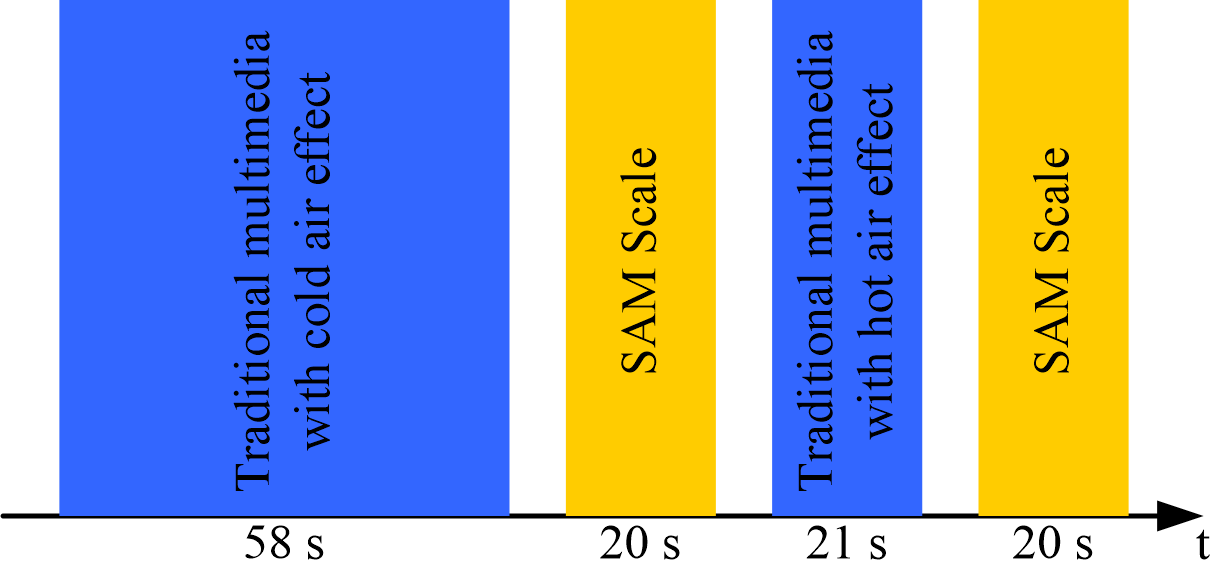}
\caption{Procedural diagram for EEG data acquisition against hot and cold air enhanced multimedia clips.}
\label{fig:procedural}
\end{center}       
\end{figure}

A Muse EEG headset was used for EEG data acquisition from $21$ participants ($11$M and $10$F). The Institution’s Ethical Review Board approved all experimental procedures involving human subjects. MUSE is a wearable headband that has four channels placed according to international 10-20 system at positions $AF7$, $AF8$, $TP9$, and $TP10$. The procedural diagram for data acquisition is shown in Fig. \ref{fig:procedural}. A clip having a duration of $58$ seconds was selected from the movie \textit{'Tangled'} and synchronized with a fan to give cold air effect. This clip was displayed to a user, whereby the EEG data was also simultaneously recorded. The 9-point self-assessment manikin (SAM) based arousal and valence scores were collected in response to the displayed clip. Another clip from the movie \textit{'The Lord of The Rings'} having a duration $21$ seconds was synchronized with a heater to give the hot air effect. The clip was displayed to the same user and in the meanwhile EEG data was recorded. The SAM based values were also gathered for the clip with hot air effect. The total procedure took $119$ seconds against each participant for viewing the clip and gathering the SAM values with an additional time of 5 minutes for briefing and data acquisition system setup. 

\subsection{Pre-processing}
The power spectrum of the EEG signals from each electrode is characterized in distinct frequency bands ranging from $1-100$ Hz. These bands are, 1$<$delta$<$4Hz, 4$<$theta$<$7Hz, 8$<$alpha$<$13Hz, 13$<$beta$<$30Hz, and gamma$>$30Hz \cite{survey19}. Different EEG bands are shown to be activated in response to different stimuli. The Muse EEG headband characterizes data in bands from each electrode. A driven right leg (DRL-REF) feedback algorithm is used in MUSE for active noise suppression. These circuits are used for proper contact of the electrodes with the head scalp. Moreover, participants were instructed to minimize their physical movements for keeping the muscular movement artifacts at a minimum.

\subsection{Feature Extraction}
Frequency domain features were used to evaluate the emotion recognition system accuracy. Three different types of frequency domain features were extracted from the pre-processed EEG data. Each feature was calculated on five bands of each channel and are described below.
\begin{enumerate}
\item Rational asymmetry (RASM) is defined as the ratio of powers on symmetrical electrode pairs. Symmetrical electrodes are the ones that are placed at the same position and distance from the center point on the right and left hemisphere of the brain, which for the MUSE headband are $(TP9,TP10)$ and $(AF7,AF8)$. The symmetry features are related to the valence of an emotional stimuli \cite{music2018}. RASM was calculated as,
\begin{equation}
RASM_B=\frac{P_{R}}{P_{L}},
\end{equation}
where $P_{R}$ and $P_{L}$ are the electrode powers on the right and left hemispheres, respectively, and $B$ represents the frequency band. A total of $10$ RASM features were calculated for five bands of the four electrodes.

\item Differential asymmetry (DASM) is the difference of powers on symmetrical electrode pairs given as,
\begin{equation}
DASM_B=P_{R}-P_{L}.
\end{equation}
A total of $10$ feature values were calculated for DASM from four electrodes. 
\item Correlation $(\rho_{a,b})$, where EEG spectral powers are correlated with different regions of the brain that directs attention \cite{2018correlation} and was computed as,

\begin{equation}
\rho_{a,b}=\frac{\Sigma(a_i-\overline{a})(b_i-\overline{b})}{\sqrt{\Sigma(a_i-\overline{a})^2\Sigma(b_i-\overline{b})^2}},
\end{equation}
where $a_i$ and $b_i$ is the $i^{th}$ value and $\overline{a}$ and $\overline{b}$ is the mean value of vectors $a$ and $b$, respectively. The feature vector length for correlation feature equals $10$.  
\end{enumerate}

We had a final feature vector of length $30$ combining all features from RASM, DASM and correlation. 

\subsection{Classification}
In the final stage, a machine learning based classification algorithm was applied for recognizing four human emotions using a support vector machine (SVM). For this purpose frequency domain features from the acquired EEG data were used. SVM has been widely for classification of human emotions using EEG signals \cite{survey}, and can be used both in classification and regression problems. SVM model takes different training instances as points in space, which are then displaced such that the sample related to different classes are separated by an apparent margin using support vectors. An SVM also maximizes the distance between hyper-planes to reduce upper bound of the generalization error. It is used for classification of non-linear and linear data by applying different kernel algorithms. An algorithm namely polynomial kernel was used in SVM in this study and is defined as,
\begin{equation}
P(a,b)=(a^{T}b+c)^{deg},
\label{eq:eq2}
\end{equation} 
where $c$ represents a constant, and $deg$ shows the polynomial degree.
%

\section{EXPERIMENTAL RESULTS}
\label{er}
\begin{figure}[t]
\begin{center}
  \includegraphics[width=85mm]{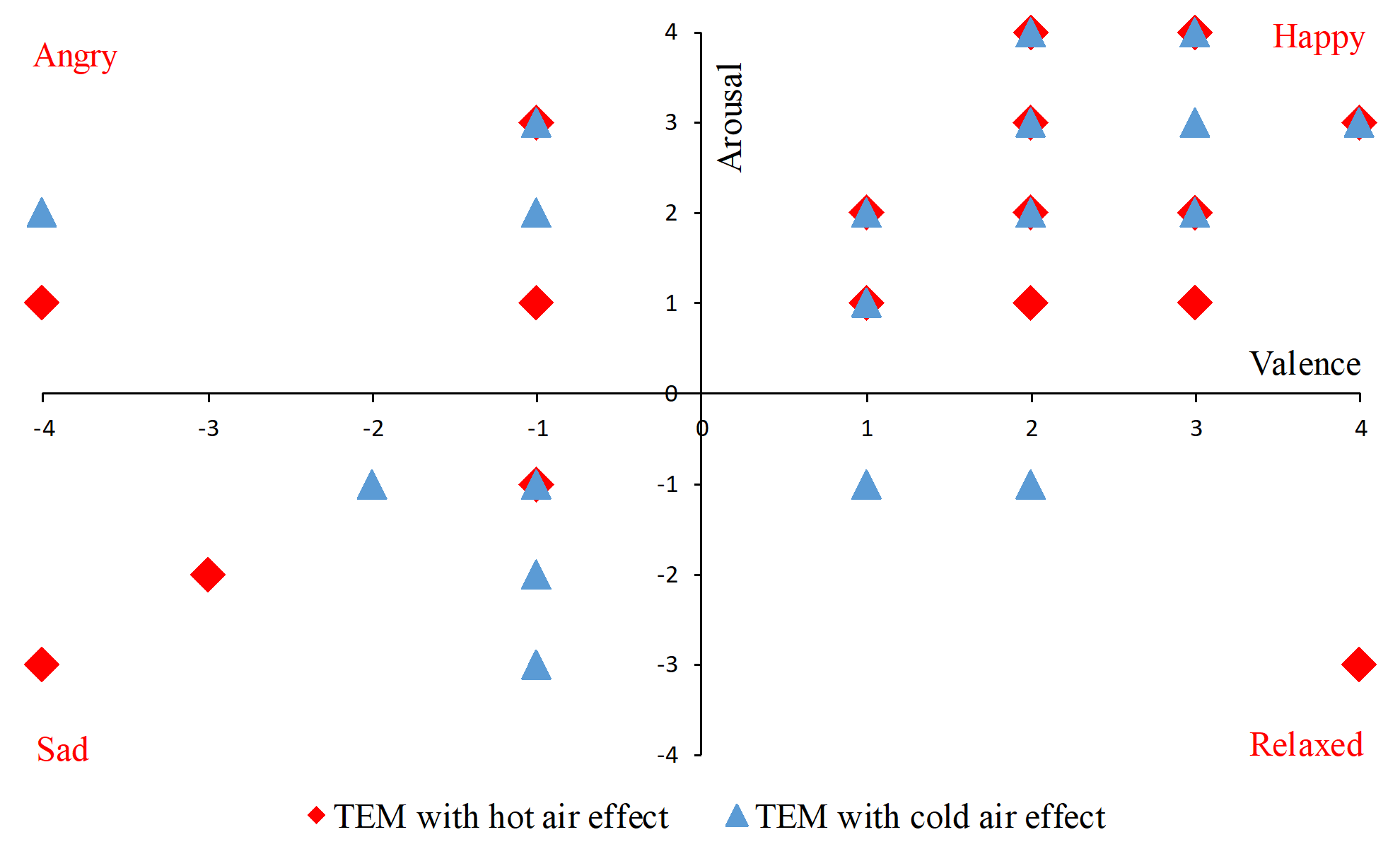}
\caption{SAM scores on a 9-point scale of arousal and valence for hot and cold air enhanced multimedia.}
\label{fig:sam}
\end{center}       
\end{figure}

The recorded data was labeled using SAM values on a 9-point scale. An emotion from each quadrant was tagged based on the SAM values \cite{samquadrant}. The emotion is happy for positive valence and arousal and sad for negative values of valence and arousal. A negative value for valence along with a positive arousal value was labeled as angry emotion, whereas a positive value of valence along with a negative value of arousal showed relaxed emotion. The valence-arousal values against all users for four labeled emotions are shown in Fig. \ref{fig:sam}. As classes are unbalanced, the re-sample algorithm was applied to balance the labeled classes. SVM was applied on each feature type separately and by fusing all features to form a larger feature vector. 

A 10-fold cross validation was applied in SVM for classification of four emotions i.e., happy, relaxed, sad, and angry by displaying TEM as a stimuli. The classifier performance was evaluated in terms of accuracy and error rates. Absolute errors i.e., mean absolute error (MAE) and root absolute error (RAE) were evaluated. Similarly squared errors i.e., root mean squared error (RMSE) and root relative squared error (RRSE) were also evaluated for SVM. Moreover, agreement of ground truth with the test data was also measured by kappa values that ranges between $-1$ to $1$, where maximum value represents a complete agreement of the ground truth with the testing data.

\begin{table}[t]
\begin{center}
\caption{Performance parameters for emotion classification using frequency domain features of brain signals}
\label{tab:tab1}       
\scalebox{0.9}{
\begin{tabular}{|c||c||c||c||c||c||c|}
\hline
Feature & Accuracy(\%) & MAE & RMSE & RAE & RRSE & Kappa\\\hline
RASM & 71.42 & 0.295 & 0.378 & 114.16 & 106.85 & 0.184 \\\hline
DASM & 66.66 & 0.299 & 0.382 & 115.70 & 108.20 & 0.050 \\\hline
Correlation & 64.28 & 0.299 & 0.383 & 115.70 & 108.32 & -0.064 \\\hline
\textbf{All} & \textbf{76.19} & \textbf{0.277} & \textbf{0.353} & \textbf{107.27} & \textbf{99.95} & \textbf{0.449}\\\hline
\end{tabular}}
\end{center}
\end{table}

\begin{table*}[!t]
\begin{center}
\caption{Performance comparison of proposed system with state-of-the-art methods for emotion recognition using EEG in response to various stimuli}
\label{tab:tab2}       
\scalebox{0.9}{
\begin{tabular}{|c||c||c||c||c||c||c||c|}
\hline
Article & Type & Features & No. of & No. of & Stimuli & No. of & Accuracy \\
& & & Electrodes & Emotions & & Participants (F/M) & \\\hline

\cite{survey112} & User Dependent & Frequency & 60 & 3 & Video & 8 (5/3) & 73\% \\\hline
\cite{survey113} & User Independent & Statistical & N/A & 7 & Video & 9 (-/-) & 36\% \\\hline
\cite{mytactile} & User Independent & Time & 4 & 4 & TEM & 21 (10/11) & 63.41\%\\\hline
Proposed & User Independent & Frequency & 4 & 4 & TEM & 21 (10/11) & 76.19\%\\\hline

\end{tabular}}
\end{center}
\end{table*}

Emotion classification results in response to TEM using frequency domain features of EEG signals are shown in Table \ref{tab:tab1}. An accuracy of $76.19$\% was achieved using EEG for four emotions by using all features with a maximum feature vector length of $30$. Whereas, RASM features gave an accuracy of $71.42$\% with the minimum feature vector length of $10$. 

\begin{figure}[t]
\begin{center}
  \includegraphics[width=90mm]{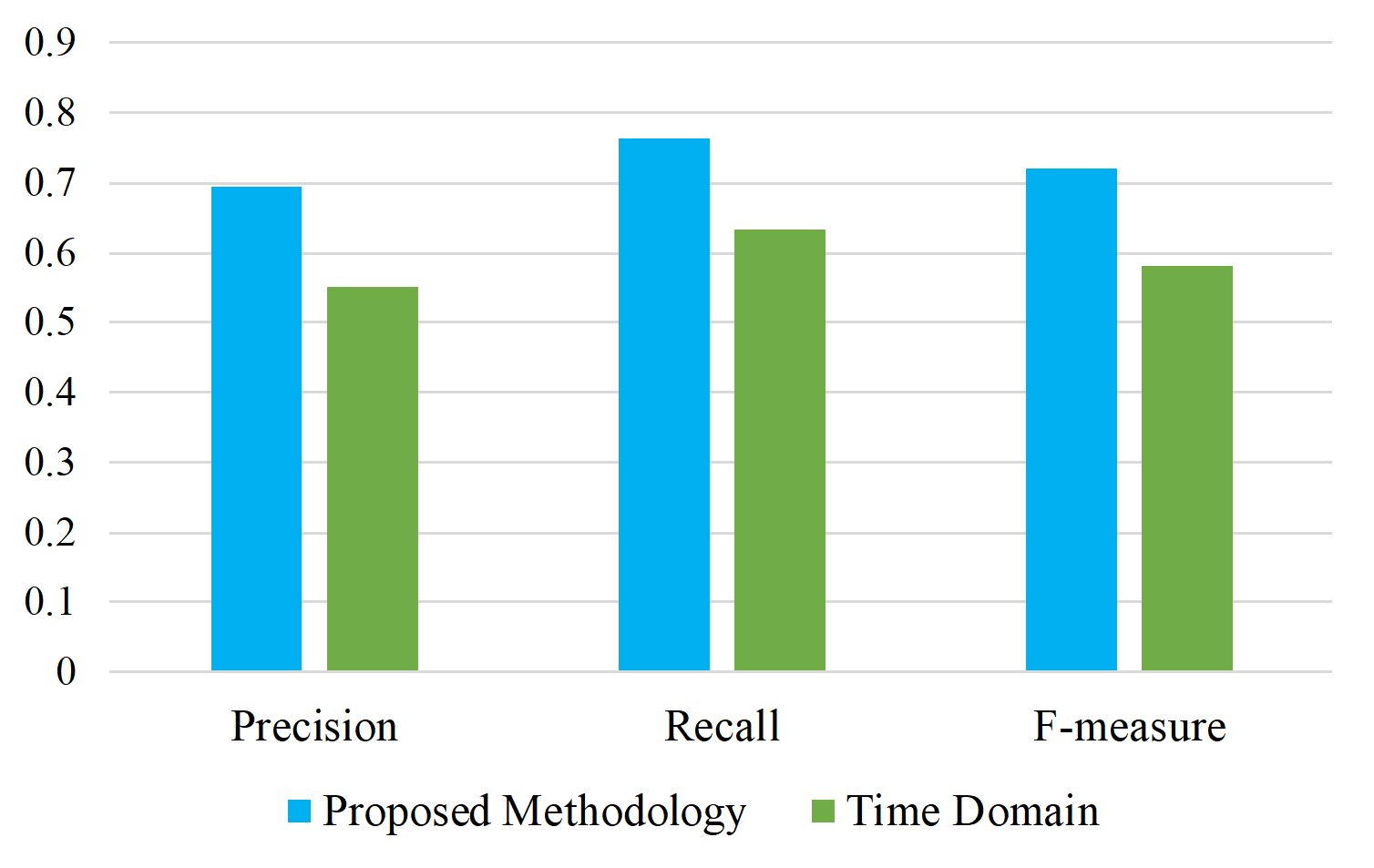}
\caption{Performance comparison of the proposed frequency domain features with time domain features \cite{mytactile} in terms of precision, F-measure, and recall.}
\label{fig:graph}
\end{center}       
\end{figure}

Recently reported studies for emotion recognition using EEG are summarized in Table \ref{tab:tab2}. These were evaluated in terms of the type, number of emotions classified, features, number of electrodes, number of subjects involved, stimuli, and accuracy. In \cite{survey112,survey113}, emotions were classified using SVM by displaying videos as a stimuli. The total number of classified emotions were $3$ and $7$, respectively. In \cite{survey112}, user dependent approach was used by extracting frequency domain features, while in \cite{survey113}, a user-independent approach was used by extracting statistical features. The accuracy and number of electrodes were 73\%(60) and 36\%(N/A) respectively. In \cite{mytactile}, a user-independent approach for four class problem against TEM reported an accuracy of $63.41$\% with four frontal electrodes using time domain features with a feature vector length of $48$. Whereas with our proposed methodology, accuracy increases to $76.19$\% by using frequency domain features for the same problem. The kappa value also increases with less error rate for four classes of emotions. 
The precision, recall, and F-measure values of the proposed method was compared with time domain statistical features against TEM in Fig. \ref{fig:graph}. It is evident that the frequency domain features has the highest precision, F-measure, and recall values as compared to time domain features against TEM.

\vspace{-2mm}
\section{CONCLUSION}
\label{cc}
Tactile enhanced multimedia content, for engaging tactile sense in addition to vision and auditory, was generated by synchronizing an electric fan and a heater with the traditional multimedia content. Brain signals were recorded using commercially available MUSE EEG headband and analyzed for four classes of emotions in response to TEM by extracting frequency domain features. The accuracy increases in comparison to time domain features. The precision, F-measure, and recall values also suggest that EEG based emotion recognition using frequency domain features in response to TEM achieves better performance in comparison to time domain features for the same content. We conclude that frequency domain EEG features could be better suited for emotion classification in response to multiple sensorial media. We intend to use multiple physiological signals in future to better understand human response to such content.      


\bibliographystyle{IEEEtran}
\bibliography{majid}

\end{document}